\documentclass[onecolumn,prc,superscriptaddress,unsortedaddress,showpacs,showkeys,preprintnumbers,amsmath,amssymb]{revtex4}

\usepackage{graphicx}
\usepackage{dcolumn}
\usepackage{bm}
\usepackage{CJK}
\usepackage{txfonts}

\def\beq{\begin{equation}}
\def\eeq{\end{equation}}
\def\bea{\begin{eqnarray}}
\def\eea{\end{eqnarray}}

\def\fun#1#2{\lower3.6pt\vbox{\baselineskip0pt\lineskip.9pt
  \ialign{$\mathsurround=0pt#1\hfil##\hfil$\crcr#2\crcr\sim\crcr}}}
 
\begin{document}
\begin{CJK*} {GBK} {song}
\preprint{}

\title{
Effects of restrained degradation on gene expression and regulation}

\author{Yan-Ling Feng}\email[Electronic address:]{ylfeng14
@lzu.edu.cn} \affiliation{The Second Hospital of Lanzhou University,
Lanzhou 730000, China}
\author{Jing Sun}
\affiliation{The Second Hospital of Lanzhou University, Lanzhou
730000, China}
\author{Yi-Fan Liu}
\affiliation{The Second Hospital of Lanzhou University, Lanzhou
730000, China}
\author{Jian-Gong Ren} \email[Electronic address:]{1310371093
@qq.com} \affiliation{The Second Hospital of Lanzhou University,
Lanzhou 730000, China}
\date{\today}
\author{Jian-Min Dong} \affiliation{Institute of Modern Physics, Chinese
Academy of Sciences, Lanzhou 730000, China}
\date{\today}

\begin{abstract}
The effects of carrying capacity of environment $K$ for degradation
(the $K$ effect for short) on the constitutive gene expression and a
simple genetic regulation system, are investigated by employing a
stochastic Langevin equation combined with the corresponding
Fokker-Planck equation for the two stochastic systems subjected to
internal and external noises. This $K$ effect characterizes the
limited degradation ability of the environment for RNA or proteins,
such as insufficient catabolic enzymes. The $K$ effect could
significantly change the distribution of mRNA copy-number in
constitutive gene expression, and interestingly, it leads to the
Fano factor slightly larger than 1 if only the internal noise
exists. Therefore, that the recent experimental measurements
suggests the Fano factor deviates from 1 slightly (Science {\bf 346}
(2014) 1533), probably originates from the $K$ effect. The $K$
effects on the steady and transient properties of genetic regulation
system, have been investigated in detail. It could enhance the mean
first passage time significantly especially when the noises are weak
and reduce the signal-to-noise ratio in stochastic resonance
substantially.

\end{abstract}


\maketitle

\section{Introduction}
Gene expression and regulations are a research focus of molecular
biology. In addition to extensive experimental studies, some
deterministic differential equations are established to
quantitatively investigate these complex processes. However, even
genetically identical cells in identical environments exhibit
variable phenotypes. This phenomenon is traced back to the gene
expression noises. The noises sometimes play crucial roles in gene
expression and
regulation~\cite{Gene1,Gene2,Gene3,Gene4,Gene5,Gene6,Gene7,Gene8,Gene9,Gene10}
that involves lots of biochemical reactions and hence many random
events. The noise can be divided into two categories: intrinsic
noise and extrinsic noise. Usually, the former one stems from the
small number of reactant molecules, while the latter one is produced
by fluctuated environment. Therefore, the deterministic approaches
that have been widely used in gene regulation networks are not valid
any longer, and hence the effect of noises should be taken into
account. Some useful approaches or equations have been proposed to
study such stochastic gene expression, such as the Gillespie
algorithms~\cite{GS1,GS2,GS3}, the master equation, the
Fokker-Planck equation and the Langevin equation, which have been
explored extensively over the past decade~\cite{FP1,FP2,FP3}.

During gene expression and regulations, the synthesized RNA or
proteins degrade persistently. The degradation rate is regarded to
be proportional to their concentration or copy number. However, due
to the limited carrying capacity of the environment for degradation,
such as the limited catabolic enzymes and ATP, the degradation rate
should be restrained and thus it deviates from this linear
relationship in particular when the concentration or copy numbers of
the RNA or proteins are large, which is analogous to the logistic
model for tumor cell growth~\cite{TM1,TM2,TM3}.

In this work, we focus on the effect of the limited carrying
capacity of the environment $K$ for degradation on the gene
expression and regulations, we call it the `$K$ effect' for the sake
of discussions. Two typical systems are investigated as illustrative
examples, that is, the constitutive gene~\cite{Gene8} and a simple
genetic regulation system given by Ref.~\cite{Sm}. For the
constitutive gene expression, it is well-known that the steady state
distribution of mRNA copy-number obeys the Poisson statistics. The
corresponding Fano factor which is defined as the ratio between the
variance $\sigma ^2$ and the mean $\overline{x}$, of the mRNA
copy-number distribution, is 1. However, the recent experimental
measurement suggests that the Fano factor for the constitutive gene
expression is larger than 1~\cite{data}, even if the contribution of
external noises is subtracted. We explore the influence of the $K$
effect on this Fano factor to attempt to explain the experimental
measurement in Sec. 2. In Sec. 3, we calculate the $K$ effect on the
simple genetic regulation model~\cite{Sm}, and investigate in detail
the steady state distribution which characterizes the steady-state
characters, the mean first passage time together with the stochastic
resonance which characterize the transient properties.

\section{Constitutive Gene Expression}
In a simplest model of constitutive gene expression, a transcript is
produced at a constant rate $\alpha$ and degraded with rate constant
$\gamma$. The kinetics is given by the deterministic differential
equation $\frac{dx}{dt}=\alpha -\gamma x$~\cite{Gene8}, where $x$ is
the copy-number of mRNA. If we include the effect of the carrying
capacity of the environment which is widely-employed in
Michaelis-Menten enzyme-catalyzed reactions, this equation should be
written as
\begin{equation}
\frac{dx}{dt}=\alpha -\gamma x(1-\frac{x}{K}),
\end{equation}
where $K$ measures the restrained degradation due to the limited
carrying capacity of the environment, which is similar with the
logistic model for tumor cell growth or population growth. With the
inclusion of both internal ($\eta (t)$) and external ($\xi(t)$)
Gaussian white noises, the corresponding Langevin equation is given
as
\begin{eqnarray}
\frac{dx}{dt} &=&\alpha -\gamma x(1-\frac{x}{K})+x(1-\frac{x}{K})\xi
(t)+\eta (t),  \nonumber \\
&<&\xi (t)>=0, \nonumber \\
&<&\xi (t)\xi (t^{\prime })>=2D\delta (t-t^{\prime }), \nonumber \\
&<&\eta (t)>=0, \nonumber  \\
&<&\eta (t)\eta (t^{\prime })>=2Q\delta (t-t^{\prime }). \label{AA}
\end{eqnarray}
The two kinds of noises from different sources are assumed to be
independent with each other. The Fokker-Planck equation
corresponding to this Langevin equation is given by
\begin{eqnarray}
\frac{\partial P(x,t)}{\partial t} &=&-\frac{\partial }{\partial x}%
A(x)P(x,t)+\frac{\partial ^{2}}{\partial x^{2}}G^{2}(x)P(x,t), \\
A(x) &=&\alpha -\gamma x(1-\frac{x}{K}), \nonumber \\
G(x) &=&\sqrt{Dx^{2}(1-\frac{x}{K})^{2}+Q}.\nonumber \label{AA1}
\end{eqnarray}
The stationary probability distribution $P(x)$ for this
Fokker-Planck equation is given by
\begin{eqnarray}
P(x)=\frac{N_0}{\sqrt{Dx^{2}(1-\frac{x}{K})^{2}+Q}}\exp \left[ \int^{x}\frac{%
\alpha -\gamma
x(1-\frac{x}{K})}{Dx^{2}(1-\frac{x}{K})^{2}+Q}dx\right],
\end{eqnarray}
where $N_0$ is a normalization constant. If only the intrinsic noise
exists, i.e., $D=0$, the probability distribution of mRNA
copy-number $x$ follows the distribution of
\begin{eqnarray}
P(x)=N\exp \left[ -\frac{(x-\frac{\alpha }{\gamma })^{2}}{2\frac{Q}{\gamma }}%
\right] \exp \left( \frac{\gamma }{3KQ}x^{3}\right),\label{BB1}
\end{eqnarray}
where $N$ is a new normalization constant. If degradation is not
restrained, i.e., carrying capacity of the environment is infinity
($K\rightarrow \infty $), the model is reduced to an unrestrained
one that has been extensively discussed previously, for which the
probability distribution of mRNA copy-number $x$ follows a Poisson
statistics~\cite{Gene8}. It is well known that the variance $\sigma
^2$ is equal to the mean $\overline{x}$ for Poisson distribution,
and hence the Fano factor defined by $\text{Fano}=\sigma
^2/\overline{x}$ is exactly 1. The Poisson distribution is
approximated to be Gauss distribution when the $\overline{x}$ is
large, and the Eq. (\ref{BB1}) is exactly a Gauss distribution in
the case of $K\rightarrow \infty $. Accordingly, the intrinsic noise
strength is $Q=\alpha$ to make sure the $\text{Fano}=1$. The limited
carrying capacity of the environment $K$, in particular when the $K$
value is small, leads to a result that the distribution deviates
from the Gauss distribution, as shown in the above Eq. (\ref{BB1}).

\begin{figure}[htbp]
\begin{center}
\includegraphics[width=0.5\textwidth]{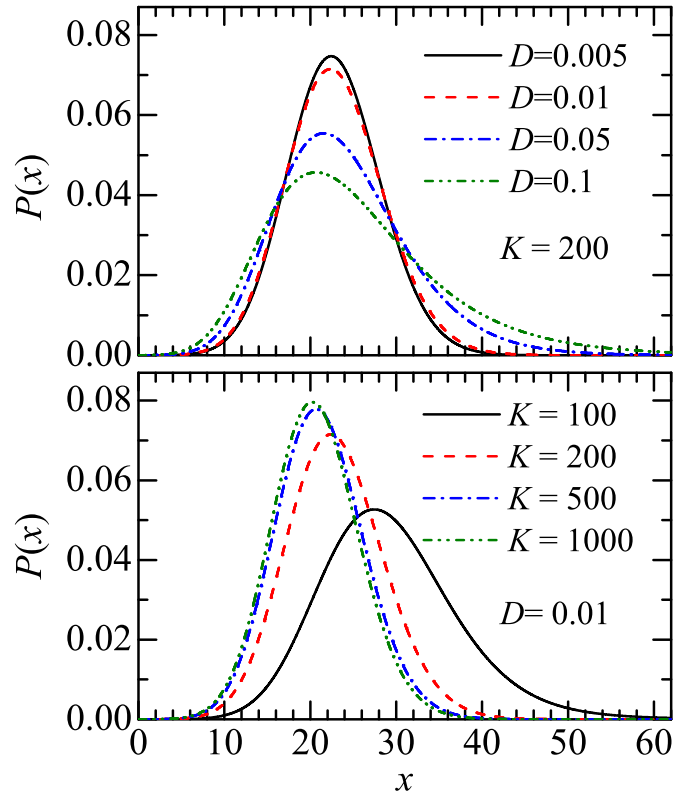}
\caption{The stationary probability distribution $P(x)$ as a
function of the mRNA copy-number $x$ under different extrinsic noise
strength $D$ and carrying capacities of the environment $K$. The
parameters we used are $\alpha=20$, $\gamma=1$, $K=200$ for upper
panel and $D=0.01$ for lower panel.}\label{fig:Pst}
\end{center}
\end{figure}

Figure~\ref{fig:Pst} displays the normalized distribution $P(x)$ as
a function of mRNA copy-number $x$ under different carrying
capacities of the environment $K$ for degradation and extrinsic
noise strength $D$. The expected single peak structure is presented.
For a given $K$, the peak value of the distribution $P(x)$ decreases
as the noise strength $D$ increases, but the shape of the $P(x)$
becomes wider and wider and more asymmetric. A larger $K$ combined
with a smaller $D$ gives a $P(x)$ much closer to a Gauss
distribution. The lower panel presents the role of the $K$ effct in
the distribution $P(x)$. Note that a small $K$ value corresponds a
weak carrying capacities of the environment, i.e., the degradation
is strongly restrained by the environment. As the the $K$ value
becomes smaller (the degradation is more strongly restrained), the
peak value of distribution $P(x)$ reduces and the location of the
peak shifts towards the larger $x$ value substantially. Similarly,
the distribution becomes more dispersive. When the $K$ value is
larger enough, for instance, $K>500$, the distribution $P(x)$ does
not change visibly because the carrying capacities of the
environment is so large that the system approaches an unrestrained
degradation case.

\begin{figure}[htbp]
\begin{center}
\includegraphics[width=0.5\textwidth]{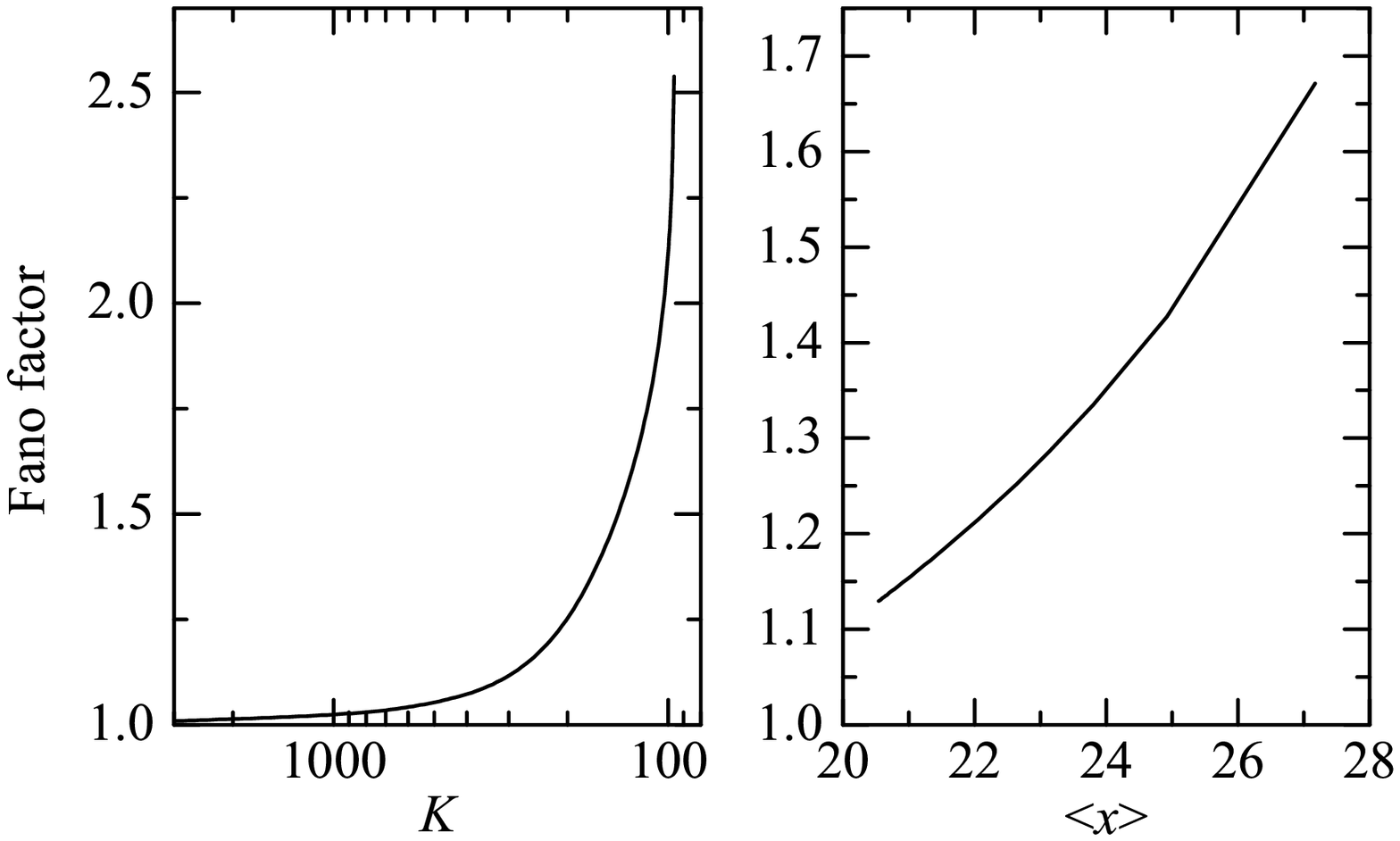}
\caption{The Fano factor versus the (a) carrying capacities of the
environment $K$ and (b) the mean copy number of mRNA $<x>$. The
parameters are $\alpha=20$, $\gamma=1$.}\label{fig:Fano}
\end{center}
\end{figure}

The Fano factor quantifies the deviation from Poisson
statistics~\cite{Gene8} for a gene expression system subject to
intrinsic noise. Figure~\ref{fig:Fano} shows the calculated Fano
factor as a function of carrying capacities of the environment $K$
and the mean mRNA copy number, where the external noise is switched
off ($D=0$) and the stationary probability distribution $P(x)$ is
described by Eq. (\ref{BB1}). The Fano factor approaches 1, when the
$K$ value becomes larger and larger and the corresponding the mean
mRNA copy number becomes larger, which is consistent with the fact
that the Fano factor for unrestrained degradation is exactly 1, and
thus it indicates the reliability of the present calculations. The
Fano factor is found to increase monotonously as the $K$ effect
becomes more stronger. Under the influence of the intrinsic noise,
the $K$ effect could induce the deviation of the Fano factor from 1,
which is different from a time delay effect~\cite{Feng2017} and is
perhaps important for us to understand gene expression. The recent
experimental measurement of the constitutive expression has
indicated that the Fano factor is slightly larger than 1 in
particular when the mean mRNA copy number per gene copy is large
(the large copy number means that the Poisson distribution
transitions to a Gauss distribution), even if the known external
noise is subtracted~\cite{data}. Therefore, the deviation of the
Fano factor from 1 is probably induced by the finite carrying
capacities of the environment for degradation.

\section{A Simple Gene Transcription Regulatory System}

Smolen {\it et al}. proposed a typical model for the gene
transcriptional regulatory system~\cite{Sm} with the deterministic
differential equation written as $\frac{dx}{dt}=\alpha +\beta
x^{2}/(x^{2}+c^{2})-\gamma x$. Here $x$ is the concentration of the
transcription factor (TF) monomer, $\alpha$ the basal synthesis
reaction rate of the TF, $\beta$ the maximal rate of phosphorylated
dimer TF activator, $\gamma$ the degradation rate, and $c$ the
concentration of $x$ needed for half-maximal induction. By
introducing the environmental fluctuation, this model has been
widely investigated due to its role of
representativeness~\cite{TF1,TF2,TF3,TF4,TF5,TF6,TF7,TF8,TF9}. Note
that the effect of a carrying capacity of the environment on the
production of the TF has been included in this model. In the present
work, we taken into account the effect of the limited carrying
capacity of the environment $K$ on the degradation rate as for the
constitutive gene expression that we discussed above. With the
inclusion of both internal ($\eta (t)$) and external ($\xi(t)$)
Gaussian white noises, the stochastic differential equation takes
the form of
\begin{eqnarray}
\frac{dx}{dt} &=&\alpha +\beta \frac{x^{2}}{x^{2}+c^{2}}-\gamma x(1-\frac{x}{%
K})+x(1-\frac{x}{K})\xi (t)+\eta (t), \nonumber  \\
&<&\xi (t)>=0, \nonumber \\
&<&\xi (t)\xi (t^{\prime })>=2D\delta (t-t^{\prime }), \nonumber \\
&<&\eta (t)>=0, \nonumber \\
&<&\eta (t)\eta (t^{\prime })>=2Q\delta (t-t^{\prime }),\nonumber \\
&<&\xi (t)\eta (t^{\prime })>=0.  \label{CC}
\end{eqnarray}
The corresponding Fokker-Planck equation is also given by Eq.
(\ref{AA1}), but with $A(x)=\alpha +\beta
\frac{x^{2}}{x^{2}+c^{2}}-\gamma x(1-\frac{x}{K})$. As a result, the
stationary probability distribution is
\begin{eqnarray}
P(x)=\frac{N}{\sqrt{Dx^{2}(1-\frac{x}{K})^{2}+Q}}\exp \left[ \int^{x}\frac{%
\alpha +\beta \frac{x^{2}}{x^{2}+c^{2}}-\gamma x(1-\frac{x}{K})}{Dx^{2}(1-%
\frac{x}{K})^{2}+Q}dx\right].
\end{eqnarray}

\begin{figure}[htbp]
\begin{center}
\includegraphics[width=0.5\textwidth]{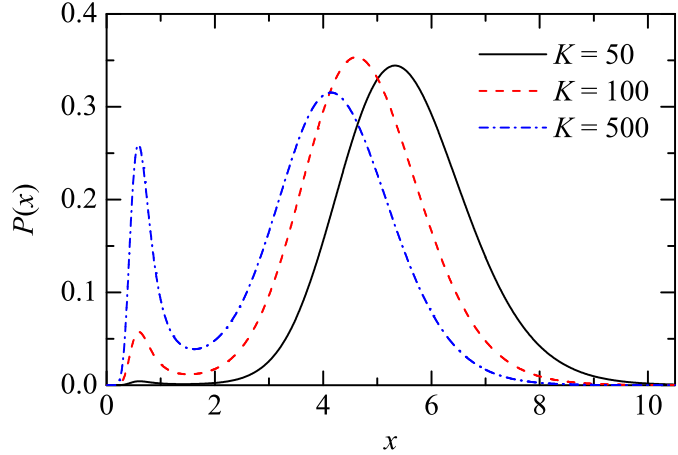}
\caption{The stationary probability distribution $P(x)$ versus the
TF monomer concentration under different carrying capacities of the
environment $K$. The parameters are $\alpha=0.4$, $\beta=6$,
$\gamma=1$, $c^2=10$, $D=0.02$, and $Q=0.005$.} \label{fig:Pst2}
\end{center}
\end{figure}

Figure~\ref{fig:Pst2} illustrates the normalized steady-state
probability distribution $P(x)$ versus the TF monomer concentration
$x$ for different different $K$ values. The parameters we used in
this work are $\alpha=0.4$, $\beta=6$, $\gamma=1$, $c^2=10$, as in
Refs.~\cite{TF2,TF3,TF4,TF8}. All the $P(x)$ present obvious bimodal
structure. One is at the low TF monomer concentration $x_1$, and the
other one is at high concentration $x_2$. Such a bistability is the
central characteristic for gene switches. When the degradation is
restrained more strongly (smaller $K$ values), the first peak of the
distribution decreases while the second one shifts towards a larger
$x$ substantially. Therefore, the $K$ effect results in the
transition from lower concentration state to a higher one.
Interestingly, when the degradation is restrained strongly by the
environment, such as $K=50$, the bimodal structure vanishes, and
just a single peak exists, as shown in Fig.~\ref{fig:Pst2}. This
indicates the importance of the $K$ effect in gene transcription
regulatory system.

\begin{figure}[htbp]
\begin{center}
\includegraphics[width=0.45\textwidth]{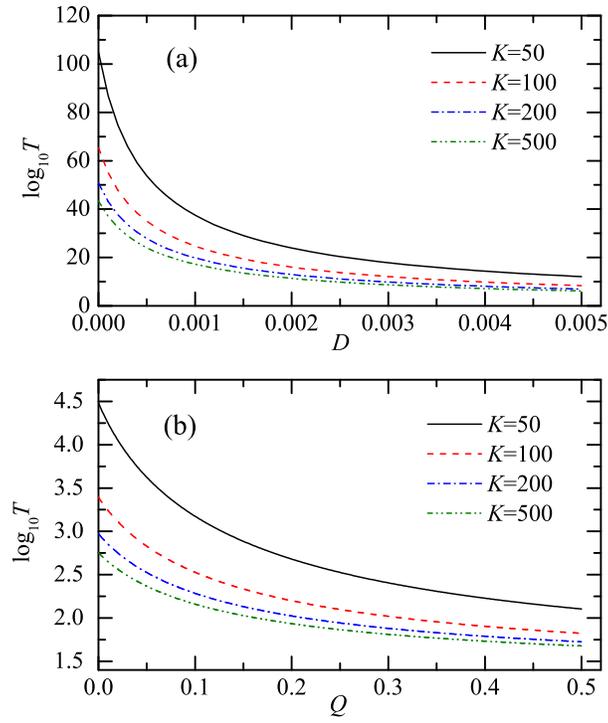}
\caption{The mean first passage time versus the noise strength $D$
and $Q$ for different carrying capacities of the environment
$K$.}\label{fig:MFPT}
\end{center}
\end{figure}

In addition to the steady state characters that we discussed above,
to characterize a stochastic dynamical system, the transient
properties are necessary. The mean first passage time is usually
used to describe the transient properties, which is defined as the
escape time from one state to the other one, and expression from the
initially state $x_2$ to final state $x_1$ is given
by~\cite{MFPT1,MFPT2}
\begin{equation}
T\simeq \frac{2\pi }{\sqrt{U_{0}^{\prime \prime
}(x_{u})U_{0}^{\prime \prime }(x_{2})}}\exp \left[ \Phi (x_{u})-\Phi
(x_{2})\right],
\end{equation}
where $U_{0}(x)$ is the deterministic potential for the system with
maximum value at $x_u$ and $\Phi
(x)=-\int^{x}\frac{A(x)}{G^{2}(x)}dx$ is the modified potential. The
mean first passage time as functions of noise strength $D$ and $Q$
for different $K$ values is presented in Fig.~\ref{fig:MFPT}. The
mean first passage time is found to monotonously decreases as the
noise strength $D$ or $Q$ increases, and decreases distinctly as the
$K$ increases in particular when the noises are weak. The decrease
of the mean first passage time means that the transition from a high
concentration state to a low one is enhanced. Therefore, a strong
$K$ effect reduces the transition between the two states obviously.

\begin{figure}[htbp]
\begin{center}
\includegraphics[width=0.45\textwidth]{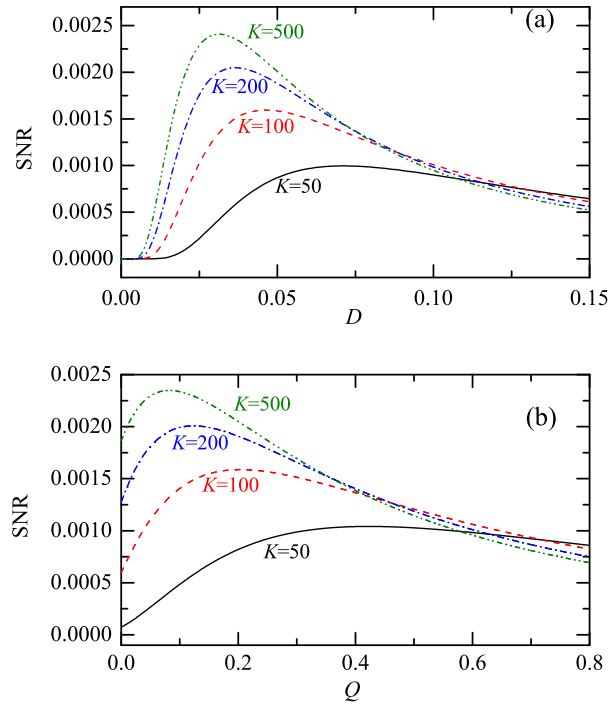}
\caption{The signal-to-noise ratio (SNR) versus the external noise
intensity $D$ (upper panel) and internal noise intensity $Q$ (lower
panel) under different carrying capacities of the environment $K$.
The intensity and frequency of the periodic signal are $A_0=0.1$ and
$\Omega=0.05$, respectively.}\label{fig:SNR}
\end{center}
\end{figure}

Stochastic resonance that was first proposed by Benzi et
al.~\cite{SR01} and Nicolis et al.~\cite{SR02} to describe the
periodic recurrences of ice ages on Earth has been studied in a
variety of nonlinear systems such as in biological systems and laser
systems~\cite{SR1,SR2,SR3,SR4,SR5,SR6,SR7}. As an intriguing
phenomenon, it is the result of the cooperative effect between the
noises and the external periodic signal. We explore the effect of
the carrying capacities of the environment $K$ for degradation on
the stochastic resonance in the gene transcription regulatory
system. The evolution of the TF concentration $x(t)$ is employing
the Langevin equation of Eq. (\ref{AA}) by adding a weak periodic
signal of $A_{0}\cos (\Omega t)\frac{x^{2}}{x^{2}+c^{2}}$, where
$A_{0}$ and $\Omega$ denote the intensity and frequency of this
signal. According to the theory proposed in Ref.~\cite{MW}, in the
adiabatic limit, the signal-to-noise ratio (SNR) is given as
\begin{equation}
\text{SNR}=\frac{\pi W_{1}^{2}A_{0}^{2}}{4W_{0}}\left[ 1-\frac{W_{1}^{2}A_{0}^{2}}{%
2\left( W_{0}^{2}+\Omega ^{2}\right) }\right] ^{-1},
\end{equation}
with
\begin{eqnarray}
W_{0} &=&2/T, \\
W_{1} &=&\left[ \Phi _{2}(x_{2})-\Phi _{2}(x_{u})\right] W_{0}, \\
\Phi _{2}(x) &=&-\int^{x}\frac{\frac{x^{2}}{x^{2}+c^{2}}dx}{Dx^{2}(1-\frac{x%
}{K})^{2}+Q}.
\end{eqnarray}
Figure~\ref{fig:SNR} shows the SNR as a function of the external
noise intensity $D$ and as a function of the internal noise
intensity $Q$ under different $K$ values. The SNR is a
non-monotonous function with respect to the noise intensity $D$ and
$Q$. The existence of maximum value in these curves is exactly the
characteristics of the stochastic resonance. When the degradation is
restrained more strongly (smaller $K$ values), the peak value of the
SNR decreases and its location shifts to the right substantially
both in the SNR-$D$ plots and in SNR-$Q$ plots, indicating the
significant role of the $K$ effect for reducing the stochastic
resonance of the gene transcription regulatory system.

\section{Summary}
We have studied the dynamics of constitutive gene expression under
the influence of carrying capacity of the environment $K$ for
degradation and noises. The distribution $P(x)$ of mRNA copy-number
has been derived by using a Fokker-Planck equation based on a
stochastic Langevin equation. A weak carrying capacity of the
environment $K$, corresponding a strong constraint on the
degradation, changes the shape of the single peak distribution of
$P(x)$ obviously. Importantly, it results in an interesting
conclusion that the Fano factor is slightly larger than 1 if only
the internal noise exists, which may be used to explain the recent
experimental measurements. The second example we discussed is the
$K$ effect on the simple genetic regulation model. The stationary
probability distribution, the mean first passage time along with the
stochastic resonance have been calculated. We found that the $K$
effect induces the transition of the probability distribution
between the high concentration state and the low state, and the
bistable structure would vanish when the $K$ effect is stronger
enough. The $K$ effect influences the mean first passage time
significantly especially when the noises are weak. The larger the
$K$ effect, the larger the mean first passage time is, and hence the
transition between the high and low concentration states is more
difficult. Finally, we found that the $K$ effect reduces the SNR in
stochastic resonance on the whole, and changes the position of the
resonance peak.

\section*{Acknowledgement}
This work is supported by the Natural Science Foundation of Gansu
Province for Distinguished Young Scholars (No. 1308RJYA065), by
Medical Health Profession Plan Project of Gansu Province (No.
GSWSKY-2015-64), by Cuiying Scientific and Technological Innovation
Program of Lanzhou University Second Hospital (No. CY2018-QN08), by
the National Natural Science Foundation of China (No. 11775276) and
by the Youth Innovation Promotion Association of Chinese Academy of
Sciences. Yan-Ling Feng and Jing Sun contributed equally to this
work.

\end{CJK*}


\begin{thebibliography}{00}

\bibitem{Gene1}
A. Becskei, L. Serrano, Nature {\bf 405}, 590 (2000).

\bibitem{Gene2}
K. Ahmad and S. Henikoff, Cell {\bf 104}, 839 (2001).

\bibitem{Gene3}
M. Elowitz, A. Levine, E. Siggia, and P. Swain, Science {\bf 297},
1183 (2002).

\bibitem{Gene4}
J. M. Pedraza and A. van Oudenaarden, Science {\bf 307}, 1965
(2005).

\bibitem{Gene5}
D. Bratsun, D. Volfson, L. S. Tsimring, and J. Hasty, PNAS {\bf
102}, 14593 (2005).

\bibitem{Gene6}
A. Sanchez and J. Kondev, PNAS {\bf 105}, 5081 (2008).

\bibitem{Gene7}
L.-h. So, A. Ghosh, C. Zong, L. A Sepulveda, R. Segev, I. Golding,
Nat. Gene. {\bf 43}, 554 (2011).

\bibitem{Gene8}
B. Munsky, G. Neuert, and A. van Oudenaarden, Science {\bf 336}, 183
(2012).

\bibitem{Gene9}
L. S. Tsimring, Rep. Prog. Phys. {\bf 77}, 026601 (2014).

\bibitem{Gene10}
G. Anandamohan, J. Biol. Phys. {\bf 41}, 49 (2015).

\bibitem{GS1}
D. T. Gillespie, J. Phys. Chem. {\bf 81}, 2340 (1977).

\bibitem{GS2}
M. A. Gibson and J. Bruck, J. Phys. Chem. {\bf 104}, 1876 (2000).

\bibitem{GS3}
D. T. Gillespie, J. Chem. Phys. {\bf 115}, 1716 (2001).

\bibitem{FP1}
T. Kepler and T. Elston, Biophys. J. {\bf 81}, 3116 (2001).

\bibitem{FP2}
T. Tian, {\it et al}., J. Comput. Appl. Math. {\bf 205}, 696 (2007).

\bibitem{FP3}
T. Bretta and T. Gallab, J. Chem. Phys. {\bf 140}, 124112 (2014).

\bibitem{TM1}
W. R. Zhong, Y. Z. Shao, Z. H. He, Phys. Rev. E {\bf 73}, 060902(R)
(2006).

\bibitem{TM2}
C. J. Wang, D. Li, D. C. Mei, Commun. Theor. Phys. {\bf 52}, 463
(2009).

\bibitem{TM3}
L. C. Du, D. C. Mei, Phys. Lett. A {\bf 374}, 3275 (2010).

\bibitem{Sm}
P. Smolen, D. A. Baxter and J. H. Byrne, Am. J. Physiol {\bf 274},
531 (1998).

\bibitem{Feng2017}
Y.-L. Feng, J.-M. Dong, D. Wang, and X.-L. Tang, Commun. Theor.
Phys. {\bf 68}, 357 (2017).

\bibitem{data}
D. L. Jones, R. C. Brewster, R. Phillips, Science {\bf 346}, 1533
(2014).

\bibitem{TF1}
Q. Liu and Y. Jia, Phys. Rev. E {\bf 70}, 041907 (2004).

\bibitem{TF2}
C. H. Zeng and X. C. Wei, Chin. Phys. Lett. {\bf 25}, 1587 (2008).

\bibitem{TF3}
C. H. Zeng and C. W. Xie, Phys. Scr. {\bf 78}, 035801 (2008).

\bibitem{TF4}
C. J. Wang, Chin. Phys. B {\bf 19}, 030503 (2010).

\bibitem{TF5}
C. Y. Bai, Y. Yan, and D. C. Mei, Chin. Phys. B {\bf 19}, 060503
(2010).

\bibitem{TF6}
C. J. Wang, Acta Phys. Sin. {\bf 61}, 010503 (2012).

\bibitem{TF7}
T. Yang, {\it et al}., J. Stat. Mech. {\bf 12}, 12015 (2014).

\bibitem{TF8}
Y.-L. Feng, J.-M. Dong, and X.-L. Tang, Chin. Phys. Lett. {\bf 33},
108701 (2016).

\bibitem{TF9}
Y. L. Feng, J. Zhu, M. Zhang, L. L. Gao, Y. F. Liu, and J. M. Dong,
Int. J. Mod. Phys. B {\bf 30}, 1650067 (2016).

\bibitem{MFPT1}
Y. Jia and J. R. Li, Phys. Rev. E {\bf 53}, 5764 (1996).

\bibitem{MFPT2}
D. C. Mei, G. Z. Xie, L. Cao, and D. J. Wu, Phys. Rev. E {\bf 59},
3880 (1999).

\bibitem{SR01}
R. Benzi, A. Sutera, and A. Vulpiani, J. Phys. A {\bf 14}, 453
(1981).

\bibitem{SR02}
C. Nicolis and G. Nicolis, Tellus {\bf 33}, 225 (1981).

\bibitem{SR1}
Y. R. Zhou, Chin. Phys. B {\bf 20}, 010501 (2011).

\bibitem{SR2}
K. K. Wang and X. B. Liu, Chin. Phys. Lett. {\bf 30}, 70504 (2013).

\bibitem{SR3}
L. F. Lin, Y. Tian, and H. Ma, Chin. Phys. B {\bf 23}, 080503
(2014).

\bibitem{SR4}
K. K. Wang and X. B. Liu, Chin. Phys. B {\bf 23}, 010502 (2014).

\bibitem{SR5}
J. H. Li, Chin. Phys. Lett. {\bf 31}, 030502 (2014).

\bibitem{SR6}
D. X. Yang, F. S. Gu, G. J. Feng, Y. M. Yang, and B. Andrew, Chin.
Phys. B {\bf 24}, 110502 (2015).

\bibitem{SR7}
Y. X. Wang, J. Q. Zhai, W. W. Xu, G. Z. Sun, and P. H. Wu, Chin.
Phys. Lett. {\bf 32}, 097401 (2015).

\bibitem{MW}
B. McNamara and K. Wiesenfeld, Phys. Rev. A {\bf 39}, 4854 (1989).

\end{thebibliography}
\end{document}